\documentclass[prd,superscriptaddress,twocolumn,nofootinbib,showpacs,preprintnumbers,floatfix]{revtex4}

\usepackage{mathrsfs}
\usepackage{amsfonts,amssymb,amsmath}
\usepackage{graphicx,epsfig}
\usepackage{multirow}
\usepackage{verbatim}

\def\fsu5{$\cal{F}$-$SU(5)$}
\def\bfsu5{$\boldsymbol{\mathcal{F}}$-$\boldsymbol{SU(5)}$}

\def\m1half{$M_{1/2}$}
\def\m3half{$M_{3/2}$}
\def\m32{$M_{32}$}

\def\mt2{$M_{T2}$}
\def\x2{$\chi^2$}
\def\2b{$M_{T2}b$}

\def\bs0{$B_S^0 \rightarrow \mu^+ \mu^-$}

\begin{document}

\title{Testing No-Scale Supergravity with the Fermi Space Telescope LAT}

\author{Tianjun Li}

\affiliation{State Key Laboratory of Theoretical Physics and Kavli Institute for Theoretical Physics China (KITPC),
Institute of Theoretical Physics, Chinese Academy of Sciences, Beijing 100190, P. R. China}

\affiliation{George P. and Cynthia W. Mitchell Institute for Fundamental Physics and Astronomy, Texas A$\&$M University, College Station, TX 77843, USA}

\author{James A. Maxin}

\affiliation{Department of Physics and Astronomy, Ball State University, Muncie, IN 47306 USA}

\author{Dimitri V. Nanopoulos}

\affiliation{George P. and Cynthia W. Mitchell Institute for Fundamental Physics and Astronomy, Texas A$\&$M University, College Station, TX 77843, USA}

\affiliation{Astroparticle Physics Group, Houston Advanced Research Center (HARC), Mitchell Campus, Woodlands, TX 77381, USA}

\affiliation{Academy of Athens, Division of Natural Sciences, 28 Panepistimiou Avenue, Athens 10679, Greece}

\author{Joel W. Walker}

\affiliation{Department of Physics, Sam Houston State University, Huntsville, TX 77341, USA}


\begin{abstract}

We describe a methodology for testing No-Scale Supergravity by the LAT instrument onboard the Fermi Space Telescope
via observation of gamma ray emissions from lightest supersymmetric (SUSY) neutralino annihilations.
For our test vehicle we engage the framework of the supersymmetric grand unified model No-Scale Flipped $SU(5)$
with extra vector-like flippon multiplets derived from F-Theory, known as \fsu5. We show that through
compression of the light stau and light bino neutralino mass difference, where internal
bremsstrahlung (IB) photons give a dominant contribution, the photon yield from annihilation of SUSY
dark matter can be elevated to a number of events potentially observable by the Fermi-LAT in the coming years.
Likewise, the increased yield in No-Scale \fsu5 may also have rendered the existing observation of a 133 GeV
monochromatic gamma ray line visible, if additional data should exclude systematic or statistical explanations.
The question of intensity aside, No-Scale \fsu5 can indeed provide a natural weakly interacting massive particle
(WIMP) candidate with a mass in the correct range to yield $\gamma \gamma$ and $\gamma Z$ emission lines
at $m_{\chi} \sim 133$ GeV and $m_{\chi} \sim 145$ GeV, respectively. Additionally, we elucidate the emerging
empirical connection between recent Planck satellite data and No-Scale Supergravity cosmological models which
mimic the Starobinsky model of inflation. Together, these experiments furnish rich alternate avenues
for testing No-Scale \fsu5, and similarly structured models, the results of which may lend
independent credence to observations made at the LHC.

\end{abstract}



\preprint{ACT-9-13, MIFPA-13-30}

\maketitle


\section{Introduction}

Supersymmetry (SUSY) provides an elegant solution to naturally resolve the gauge hierarchy problem
within the Standard Model (SM), and presuming $R$ parity conservation, the lightest supersymmetric
particle (LSP) neutralino serves as a viable cold dark matter (CDM)
candidate~\cite{Goldberg:1983nd,Ellis:1983ew}. The empirical search for a weakly interacting
massive particle (WIMP) currently evolves on multiple fronts. For instance, the
Large Hadron Collider (LHC) at CERN sifts through trillions of proton-proton collisions for a rare
glimpse of an anomalous missing transverse energy component of hypothetical supersymmetric
interactions, where the SUSY LSP escapes the detector without direct observation as a consequence of its
neutral $U(1)_{EM}$ charge and status as an $SU(3)_C$ singlet. Sharing an equivalent objective, the
XENON~\cite{Aprile:2012nq}, CDMS~\cite{Agnese:2013rvf}, and LUX~\cite{Akerib:2013tjd}
experiments parse through statistics gathered from ionization and scintillation of inert gases and
semiconductors to potentially uncover direct observation of elastic collisions of a WIMP within the
scintillating material. Likewise, the
Fermi Space Telescope~\cite{Atwood:2009ez} strives toward this goal through latent
observation of photon decay relics from WIMP annihilations. Status of observability for this
latter conjectural phenomena, primarily within the context of a well defined model named No-Scale \fsu5,
presides as the motivating intent of this work; this approach offers a viable link between SUSY
bino dark matter and a recently observed marginal sharp line spectra, and perhaps more pertinently,
crafts a roadmap for future discovery of bino dark matter utilizing current and forthcoming sky-scanning surveys.

The annihilation of WIMPS within inner galactic regions can be prospective sources of gamma ray
emissions that compete with the astrophysical background. SUSY LSP neutralinos can annihilate directly to
gamma rays mono-energetically, yielding a (quasi-) monochromatic energy spectrum via
annihilation processes $\widetilde{\chi} \widetilde{\chi} \to \gamma \gamma$ ($E_{\gamma} = m_{\chi}$),
$\widetilde{\chi} \widetilde{\chi} \to \gamma Z$, and $\widetilde{\chi} \widetilde{\chi} \to \gamma h$.
These processes occur at 1-loop, since WIMPS cannot directly couple to the
photons, thereby suppressing the cross-section of thermally produced dark matter. Internal
bremsstrahlung (IB) photons can also produce sharp spectral features with annihilation into
charged particles via $\widetilde{\chi} \widetilde{\chi} \to f \overline{f} \gamma$, with the benefit
that IB processes occur at tree level, thus providing a larger annihilation rate for bino neutralinos and
amplifying observability.

In 2012, a tentative 130 GeV monochromatic gamma ray line was
observed~\cite{Bringmann:2012vr,Weniger:2012tx}
in the Fermi-LAT all sky surveys, exhibiting a local signal significance of
4.3--4.6$\sigma$ (3.1--3.3$\sigma$ global). Post reprocessing of the data by the Fermi
Collaboration, the budding signal shifted closer to 133 GeV with a diminished local signal significance
of 3.3$\sigma$ (global 1.6$\sigma$)~\cite{FERMI-LAT:2013uma}, somewhat dampening the enthusiasm
for a prospective indirect discovery of dark matter. Additionally, a deviation at this same $E_{\gamma} \sim 133$
GeV has been observed by the LAT instrument in a control sample of gamma rays from the
Earth's limb, elevating the likelihood that the reported effects are systematic in origin.
Therefore, the jury remains out on the validity of the signal, and a conclusive judgment may not be
pronounced for as much as two additional years, pending additional data acquisition and analysis.
Yet, this tentative observation highlights the importance of a model dependent analysis of the Fermi-LAT's
reach into the supersymmetric parameter space. Due to the small bino annihilation cross-section of
$\langle \sigma v \rangle _{\gamma \gamma} \sim 10^{-30}~{\rm cm^3/sec}$, in comparison to the best fit of the deviation in
the Fermi-LAT data of
$\langle \sigma v \rangle _{\gamma \gamma} \sim 10^{-27}~{\rm cm^3/sec}$~\cite{Bringmann:2012vr,Weniger:2012tx},
the supersymmetric origins of the 130 GeV
monochromatic gamma ray signal were quickly dismissed~\cite{Cohen:2012me}. Minus the presence of an
extraordinarily large boost factor $(BF)$ of $BF \sim 1000$, the cross-section of the observed 130 GeV
signal seemed far too large for bino dark matter annihilations to two gamma rays to be a serious candidate.

Despite these objections to solicitation of a supersymmetric explanation for the 133 GeV gamma ray line, it
was shown that a WIMP mass capable of producing $\gamma \gamma$ emission at 133 GeV and $\gamma Z$ emission
at $\sim 145$ GeV can be naturally explained~\cite{Li:2012jf} in the supersymmetric grand unified
theory (GUT) model No-Scale flipped $SU(5)$ with extra vector-like matter multiplets called ``{\it
flippons}''~\cite{
Li:2010ws, Li:2010mi,Li:2010uu,Li:2011dw, Li:2011hr, Maxin:2011hy,
Li:2011xu, Li:2011in,Li:2011gh,Li:2011rp,Li:2011fu,Li:2011ex,Li:2011av,
Li:2011ab,Li:2012hm,Li:2012tr,Li:2012ix,Li:2012yd,Li:2012qv,Li:2012jf,Li:2012mr,Li:2013hpa,Li:2013naa,Li:2013bxh},
the model referred to as \fsu5. When considering a dominant contribution from the IB final states, the
No-Scale \fsu5 upper 2$\sigma$ limit on the WIMP mass for the observed monochromatic gamma ray line is
about $M_{1/2} \sim$ 775--800 GeV. While this particular SUSY mass is currently experiencing some
tension from the LHC SUSY search~\cite{ATLAS-CONF-2013-061}, sufficient uncertainty remains in
our spectrum calculations and Monte-Carlo simulations to likewise caution against its definitive
exclusion until the 13 TeV LHC energizes in 2015.

Nonetheless, the more pressing question facing association of
this result, or the prospect of a feasible near-term future Fermi-LAT observation at some heavier energy scale,
with a genuine SUSY signal is that of whether an abnormally large boost factor is necessary to generate the
observed photon flux. Moreover, this must be accomplished without overboosting fermion channels in the
continuum; for instance, the stau-mediated channel with a
$\widetilde{\chi} \widetilde{\chi} \to \tau^+ \tau^-$ final state, where the latest upper limit on the
annihilation cross-section from observation of cosmic rays is $\langle \sigma v \rangle _{\tau \tau}
\lesssim 5 \times 10^{-25}~{\rm cm^3/sec}$~\cite{Bergstrom:2013jra}, though parallel studies
suggest the current limit could be as low as $\langle \sigma v \rangle _{\tau \tau} \lesssim 3 \times
10^{-26} - 10^{-25}~{\rm cm^3/sec}$~\cite{Egorov:2013exa}. However, the SUSY mass $M_{1/2} \sim 775$~GeV
in No-Scale \fsu5 has an annihilation cross-section of $\langle \sigma v \rangle _{\tau \tau} = 6.8
\times 10^{-29}~{\rm cm^3/sec}$, placing the necessarily required large boost factor of $BF \sim 1000$
near the fringe of the upper limit allowed on any extraneous boost in the cross-section. It is thus
preferable to pursue another course that does not involve relying upon such a large boost factor.
Because the IB photon flux is about an order or magnitude (to be concrete, around $\sim 8-12\times$ [see Table~\ref{tab:flux}])
larger than the gamma gamma flux, the IB cross section is roughly 20 times
larger than the gamma gamma cross section. Therefore, the No-Scale \fsu5 IB cross section is on the order of 
$5 \times 10^{-29}~{\rm cm^3/sec}$.  Thus, the corresponding boost factor that is needed to explain the 133
GeV Fermi-LAT gamma ray line in this scenario is substantively smaller, on the order of 50 to 100.
If one allows that the dark matter density, which enters into the pairwise interaction as a square,
is seven to ten times larger than what is traditionally used in the dark matter subhalo, this mechanism can
explain the observed gamma ray line.

Regardless of whether the existing marginal 133 GeV gamma ray line eventually is shown to be a systematic
or statistical effect, upcoming data from the Fermi Space Telescope (or future projects including Gamma-400,
DAMPE and HERD) may provide exclusive insights into the SUSY parameter space in the No-Scale \fsu5 model.
A central task confronted by this document is classification of the gamma ray signatures associable
with \fsu5, and quantification of their detection prospects across the model space, especially in the
context of an additional six years of data collection by the Fermi-LAT instrument.
Given the reality (failing upward revisions in estimates of the dark matter density profile) that the
present generation gamma telescope will not achieve the sensitivity required to
observe bino dark matter at annihilation cross-sections of
$\langle \sigma v \rangle _{\gamma \gamma} \sim 10^{-30}~{\rm cm^3/sec}$,
we highlight a phenomenologically viable scenario where the probability of uncovering an
observable indirect detection signature is somewhat more appreciable; in particular, we shall consider increasing
the photon yield from annihilation via compression of the lightest slepton and LSP neutralino mass difference to near
degeneracy, thereby establishing upward pressure on the annihilation rate, which can further elevate
the advantage of the already dominant tree level IB effects over monochromatic loop level dark matter
annihilation. This methodology can be quite naturally accommodated
in No-Scale \fsu5 with no effect on the spectrum calculations and experimental constraints established
in the model space~\cite{Li:2013naa}. The one unavoidable consequence of such a maneuver manifests
itself in a suppressed bino neutralino relic density for $M_{1/2} \lesssim1500$ GeV, transitioning to
below the recent Planck measurements~\cite{Ade:2013zuv}, thereby compelling a non-thermal mechanism
to generate the correct dark matter density. When the mass difference between the LSP neutralino and
light stau is small, the LSP--light stau coannihilation cross section will be large, resulting in a dark
matter relic density that is smaller than the observed value. Interestingly, cosmologically late decay
of string-theoretic moduli fields provide an alternate mechanism for generating the correct dark matter
relic density~\cite{Moroi:1999zb}. As the gaugino mass is increased
from smaller values of $M_{1/2}$ in No-Scale \fsu5, a naturally occurring linear compression in the
light stau and LSP mass difference counteracts this bino relic density suppression
in~\fsu5~\cite{Li:2013naa} ({\it i.e.} elevation in the annihilation rate induced by mass degeneracy
is counteracted by simple mass suppression), eventually generating the Planck measured CDM relic density
$\Omega h^2 = 0.1199 \pm 0.0027$~\cite{Ade:2013zuv} at
$M_{1/2} \sim 1500$ GeV for a nearly degenerate light stau and LSP
($\Delta M(\widetilde{\chi}_1^0,~\widetilde{\tau}_1) \simeq 2$ GeV).

The No-Scale \fsu5 framework suggested here as a vehicle for interpreting Fermi-LAT observations has
already been well developed. The model is based upon the tripodal foundations of the dynamically
established boundary conditions of No-Scale Supergravity, the Flipped $SU(5)$ Grand
Unified Theory (GUT), and the pair of TeV-scale hypothetical flippon vector-like
super-multiplets~\cite{
Li:2010ws, Li:2010mi,Li:2010uu,Li:2011dw, Li:2011hr, Maxin:2011hy,
Li:2011xu, Li:2011in,Li:2011gh,Li:2011rp,Li:2011fu,Li:2011ex,Li:2011av,
Li:2011ab,Li:2012hm,Li:2012tr,Li:2012ix,Li:2012yd,Li:2012qv,Li:2012jf,Li:2012mr,Li:2013hpa,Li:2013naa,Li:2013bxh}
derived within local F-theory model building. The convergence of these features has been shown to
naturally resolve many longstanding theoretical issues, whilst comparing positively with real world
experimental observation. Moreover, a recent
analysis~\cite{Ellis:2013xoa,Ellis:2013nxa,Ellis:2013nka}
suggests that a cosmological model based upon the No-Scale supergravity sector yields compatibility
with the Planck satellite measurements. With convenient superpotential parameter choices, the new
cosmological model compatible with Planck data is a No-Scale supergravity realization of the
Starobinsky model of inflation~\cite{Starobinsky:1980te,Mukhanov:1981xt,Starobinsky:1983zz}.
This prospective empirical evidence of the existence of a ubiquitous No-Scale supergravity sector
amplifies our motivation for implementing No-Scale \fsu5 as a realistic framework appropriate for
evaluation against formerly recorded and forthcoming Fermi-LAT gamma ray emission statistics.

The structure of this paper is as follows. First we provide a brief review of the No-Scale \fsu5 model,
and then elaborate the interesting empirical correlation between recent Planck satellite data and cosmological
models based upon No-Scale Supergravity that realize inflation in the Starobinsky mode. Next we shall
present more detailed aspects of the IB effects on the annihilation rate and, finally, we present some
benchmark models with SUSY spectra linked to neutralino annihilation cross-sections testable by
the Fermi Space Telescope in the upcoming years, as well as benchmarks consistent with a No-Scale \fsu5 explanation of the
observed 133 GeV monochromatic gamma ray line.

\section{The No-Scale \fsu5 Model}

Mass degeneracy of the superpartners has not been observed, indicating that SUSY breaking occurs near the
TEV scale. Supergravity models are GUTs with gravity mediated supersymmetry breaking, where we can fully
characterize the supersymmetry breaking soft terms by a limited set of universal parameters: universal
gaugino mass $M_{1/2}$, universal scalar mass $M_0$, Higgsino mixing $\mu$-parameter, Higgs bilinear
$B_{\mu}$-parameter, and universal trilinear coupling $A_0$. The $B_{\mu}$ and $|\mu|$ parameters are
then determined at low energy through minimization of the Higgs potential triggering radiative
electroweak symmetry breaking (REWSB), with the sign of $\mu$ remaining undetermined. Equivalently, we
can trade $B_{\mu}$ at low energy for the low energy ratio of the Higgs vacuum expectation values (VEVs)
$\tan\beta$. Subsequently remaining are the high-energy boundary conditions $M_{1/2}$, $M_0$,
$B_{\mu}$, $A_0$, and the low energy boundary condition $\tan\beta$, plus the undetermined sign of
$\mu$, which we always take to be sgn$(\mu) > 0$, as suggested by the results of $(g_{\mu}-2)/2$ of the muon.

In order to address the cosmological flatness problem,
No-Scale Supergravity was proposed~\cite{Cremmer:1983bf} as the subspace of supergravity models
which fulfill three constraints:
i) the vacuum energy vanishes automatically due to the appropriate K\"ahler potential; ii) there exist
flat directions that leave the gravitino mass $M_{3/2}$ undetermined at the minimum of the scalar
potential; iii) the quantity ${\rm Str} {\cal M}^2$ is zero at the minimum. Large one-loop corrections
would force $M_{3/2}$ to be either identically zero or of the Planck scale if the third condition were
violated. A minimal K\"ahler potential that meets the first two conditions
is~\cite{Ellis:1984bm,Cremmer:1983bf}
\begin{eqnarray} 
K &=& -3 {\rm ln}( T+\overline{T}-\sum_i \overline{\Phi}_i
\Phi_i)~,~
\label{NS-Kahler}
\end{eqnarray}
where $T$ is a modulus field and $\Phi_i$ are matter fields, which parameterize the non-compact
$SU(N,1)/SU(N) \times U(1)$ coset space. The third condition can always be satisfied in principle and is
model dependent~\cite{Ferrara:1994kg}. From the K\"ahler potential in Eq.~(\ref{NS-Kahler}) we
automatically attain the No-Scale boundary condition $M_0 = A_0 = B_{\mu} = 0$, while $M_{1/2}$ is allowed
to be non-zero and hence evolve naturally, and in fact, is necessary for SUSY breaking. Moreover, the
high-energy boundary condition $B_{\mu} = 0$ in principle determines $\tan\beta$ at low energy. The
gravitino mass $M_{3/2}$ is determined by the equation $d(V_{EW})_{min}/dM_{3/2}=0$ due to the fact
that the minimum of the electroweak (EW) Higgs potential $(V_{EW})_{min}$ depends on $M_{3/2}$, and
consequently, the supersymmetry breaking scale is determined dynamically. We are thus left
with a natural $one$-$parameter$ model, with the sole degree of freedom being the gaugino mass $M_{1/2}$.
As a deep fundamental correlation to string theory, No-scale supergravity can be realized in the
compactification of the weakly coupled heterotic string theory~\cite{Witten:1985xb}, as well as the
compactification of M-theory on $S^1/Z_2$ at the leading order~\cite{Li:1997sk}.

Precise string-scale gauge coupling unification while also evading the Landau pole problem can be realized by
supplementing the standard ${\cal F}$-lipped
$SU(5)\times U(1)_X$~\cite{Nanopoulos:2002qk,Barr:1981qv,Derendinger:1983aj,Antoniadis:1987dx}
SUSY field content with the following TeV-scale vector-like multiplets
(flippons)~\cite{Jiang:2006hf}
\begin{eqnarray}
\hspace{-.3in}
& \left( {XF}_{\mathbf{(10,1)}} \equiv (XQ,XD^c,XN^c),~{\overline{XF}}_{\mathbf{({\overline{10}},-1)}} \right)\, ,&
\nonumber \\
\hspace{-.3in}
& \left( {Xl}_{\mathbf{(1, -5)}},~{\overline{Xl}}_{\mathbf{(1, 5)}}\equiv XE^c \right)\, ,&
\label{z1z2}
\end{eqnarray}
where $XQ$, $XD^c$, $XE^c$, $XN^c$ have the same quantum numbers as the
quark doublet, the right-handed down-type quark, charged lepton, and
neutrino, respectively. Models of this nature can be realized in ${\cal F}$-ree ${\cal F}$-ermionic string
constructions~\cite{Lopez:1992kg} and ${\cal F}$-theory model
building~\cite{Jiang:2009zza,Jiang:2009za}, and have been appropriately designated ${\cal F}$-$SU(5)$~\cite{Jiang:2009zza}.

The split-unification framework of
\fsu5~\cite{Nanopoulos:2002qk,Barr:1981qv,Derendinger:1983aj,Antoniadis:1987dx}
provides for fundamental GUT scale Higgs representations (not adjoints), natural doublet-triplet
splitting, suppression of dimension-five proton decay~\cite{Antoniadis:1987dx,Harnik:2004yp},
and a two-step see-saw mechanism for neutrino masses~\cite{Ellis:1992nq,Ellis:1993ks}.
Adjustments to the one-loop gauge $\beta$-function coefficients $b_i$ induced by inclusion of the
vector-like flippon multiplets generate the required flattening of the $SU(3)$ Renormalization Group
Equation (RGE) running ($b_3 = 0$)~\cite{Li:2010ws}, which manifests as a wide separation between the primary
$SU(3)_C \times SU(2)_L$ unification near $10^{16}$~GeV and the secondary $SU(5) \times U(1)_X$ unification
near the Planck mass. The corresponding baseline extension for logarithmic running of the
No-Scale boundary conditions, especially that of $B_\mu = 0$, permits ample scale for natural dynamic
evolution into phenomenologically favorable values consistent with experiment at the EW scale. The
$SU(3)_C$ gaugino mass scale flattening generates a stable characteristic mass texture of
$M(\widetilde{t}_1) < M(\widetilde{g}) < M(\widetilde{q})$, engendering a light stop and
gluino that are lighter than all other squarks~\cite{Li:2010ws}.

The No-Scale \fsu5 model space satisfies a minimal set of necessary constraints from theory and
phenomenology~\cite{Li:2011xu,Li:2013naa}. The constraints are: i) consistency with the
dynamically established boundary conditions of No-Scale supergravity (most significantly the strict
enforcement of a vanishing $B_{\mu}$ parameter at the ultimate flipped $SU(5)$ GUT unification near
$M_{\rm Pl}$, imposed as $\left|B_{\mu}\right(M_{\cal F})| \leq 1$ GeV, about the scale of the EW
radiative corrections); ii) REWSB; iii) the centrally observed Planck CDM relic density
$\Omega h^2 = 0.1199 \pm 0.0027$~\cite{Ade:2013zuv}) ; iv) the world average top-quark mass
$m_t = 173.3 \pm 1.1$~GeV~\cite{:1900yx}; v) precision LEP constraints on the light SUSY chargino and neutralino mass
content~\cite{LEP}; and vi) production of a lightest CP-even Higgs boson mass of $m_{h} = 125.5 \pm 1.5$
GeV, accomplished through additional tree level and one-loop contributions to the Higgs boson mass by
the flippon supermultiplets~\cite{Li:2011ab,Li:2012jf,Li:2013naa}, supplementing the Minimal
Supersymmetric Standard Model (MSSM) Higgs boson mass by just the essential additional 3-5 GeV amount
requisite to attain $m_{h} \sim 125$ GeV, while also preserving a testably light SUSY spectrum that does
not reintroduce the gauge hierarchy problem via very heavy scalars that SUSY was originally intended to
solve in the first place. A two-dimensional parameterization in the vector-like flippon super-multiplet mass scale $M_V$ and the
universal gaugino boundary mass scale $M_{1/2}$ is excised from a larger four-dimensional hyper-volume
that also includes the top quark mass $m_t$ and the ratio $\tan \beta$. The enduring model space after
application of these minimal constraints is capable of maintaining the delicate balance needed to
realize the two conditions $B_\mu = 0$ and $\Omega h^2 = 0.1199 \pm 0.0027$.

The surviving No-Scale \fsu5 model space after direct application of the constraints noted above
consists of a diagonal wedge ({\it cf.} Ref.~\cite{Li:2013naa}) in the ($M_{1/2}$, $M_V$) space, the
width of which at small $M_{1/2}$ and small $M_V$ is bounded by the LEP constraints and by the CDM constraints and the
transition to a charged stau LSP at large $M_{1/2}$ and large $M_V$. Conversely, the upper limit at large
$M_V$ and the lower limit at small $M_V$ are constrained by the central experimental range on the top quark
mass. The intersection of all constraints yields a net experimentally viable model space extending
from $M_{1/2} \simeq 400$ GeV to $M_{1/2} \simeq 1500$ GeV, with an associated vector-like flippon mass
of $M_V \simeq 1$ TeV to $M_V \simeq 180$ TeV.

\section{No-Scale Supergravity Inflation}

The elegantly minimalistic formalism of No-Scale
Supergravity~\cite{Cremmer:1983bf,Ellis:1983sf, Ellis:1983ei, Ellis:1984bm, Lahanas:1986uc}
allows for a deep fundamental correlation to string theory in the infrared limit,
the natural inclusion of general coordinate invariance (general relativity),
a supersymmetry breaking mechanism that preserves a vanishing cosmological constant at tree level
(facilitating the observed longevity and cosmological flatness of our Universe~\cite{Cremmer:1983bf}),
natural suppression of CP violation and flavor-changing neutral currents, dynamic stabilization
of the compactified spacetime by minimization of the loop-corrected scalar potential, and a powerful
contraction in parameterization freedom.

Recently, an added phenomenological boost has been given to No-Scale Supergravities by detailed
measurement of the Cosmic Microwave Background (CMB) perturbations (the structural seeds of galactic
supercluster formation residually imprinted upon the faint afterglow of the big bang) from the
Planck~\cite{Ade:2013uln} satellite.
Many important features predicted qualitatively by the cosmological inflationary paradigm have been
borne out, for instance, there are no significant signs of non-Gaussian
fluctuations or hints of non-trivial topological features such as cosmic strings.
Additionally, these observations verified a highly statistically significant tilt $n_s \simeq 0.960 \pm 0.007$
in the spectrum of scalar perturbations,
as expected if the effective scalar energy density decreased gradually during inflation,
and set stronger upper limits on the ratio $r < 0.08$ of tensor (directional) to
scalar (isotropic) perturbations. These measurements, particularly of $n_s$, place many leading
models of cosmic inflation in jeopardy (cf. Fig.~1 of Ref.~\cite{Ade:2013uln}),
although a curious scenario suggested by Starobinsky~\cite{Starobinsky:1980te}
in 1980 is known~\cite{Mukhanov:1981xt} to match the data effortlessly. This model is a rather
ad-hoc modification of Einstein's description of gravity, which combines a quadratic power of the
Ricci scalar with the standard linear term. At face value, this $(R+R^2)$ model is rather difficult to
take seriously, but there is substantial enthusiasm for the observation by
John Ellis, Keith Olive and one of the authors (D.V.N), that this esoteric model is
in fact conformally equivalent to No-Scale supergravity
with an $SU(2,1)/SU(2) \times U(1)$ K\"ahler potential~\cite{Ellis:2013xoa,Ellis:2013nxa,Ellis:2013nka}, which is a subcase of
Eq.~(\ref{NS-Kahler}). To be specific, the algebraic equations of motion corresponding
to an auxiliary scalar field $\Phi$ with a quadratic potential that couples to a conventional Einstein term
may be freely substituted back into the action, resulting in the phenomenologically favorable
quadratic power of the scalar curvature~\cite{Stelle:1977ry,Whitt:1984pd}.
In short, inflation in our \fsu5 No-Scale $SU(N,1)$ framework can be realized naturally
and is consistent with the Planck results.

\section{Testing No-Scale \fsu5 with Fermi-LAT}

The monochromatic line signals are not the only mechanism capable of generating gammas visible to the 
Fermi-LAT instrument. In fact, dark matter annihilation into two Standard Model particles with a
radiated photon, a process known as internal bremsstrahlung, can also give sharp spectral features in the
ray spectrum close to the dark matter mass~\cite{Bringmann:2007nk}. The photon can arise from the final
state radiation (FSR) or virtual charged particle radiation/virtual IB (VIB). Thus, the IB photons will
be the total contributions from both FSR and VIB.

\begin{table*}[htp]
	\centering
	\caption{Ten No-Scale \fsu5 benchmarks, with points that can satisfy the Planck satellite relic
density measurements, points with $\Delta M(\widetilde{\chi}_1^0,\widetilde{\tau}_1) \simeq 2$ GeV, and
points imposing degeneracy, $\Delta M(\widetilde{\chi}_1^0,\widetilde{\tau}_1) \simeq 0$ GeV,
between the light stau and LSP mass in order to increase the annihilation rate and raise the IB contributions.
Given are the gaugino mass $M_{1/2}$, flippon mass $M_V$, $\tan\beta$, top quark mass $m_t$, relic
density $\Omega h^2$, EM $f \overline{f}$, $\gamma \gamma$, and $\gamma Z$ annihilation cross-sections,
SUSY masses, and light Higgs boson mass $m_h$. All benchmark LSP compositions are greater than 99\%
bino. The $\Omega h^2$ shown is the thermal neutralino density calculated with {\tt MicrOMEGAs~2.4}. For
those benchmarks with $\Omega h^2 < 0.1199 \pm 0.0027$, the Planck satellite measured relic density can be
generated via non-thermal mechanisms. The annihilation cross-sections given here are the average
between the {\tt MicrOMEGAs~2.4} and {\tt DarkSUSY~5.1.1} calculations. The total
$\langle \sigma v \rangle _{f \overline{f}}$ annihilation cross-section is composed of $\langle \sigma v \rangle _{f
\overline{f}} = \langle \sigma v \rangle _{\tau^+ \tau^-} + \langle \sigma v \rangle _{t\overline{t}} +
\langle \sigma v \rangle _{b \overline{b}}$. The $\Delta M$ value refers to the lightest bino neutralino
and light stau mass difference. The light
Higgs boson mass includes both the tree level+1-loop+2-loop+3-loop+4-loop and flippon contributions.
All masses are in GeV and all cross-sections in ${\rm cm^3/sec}$.}
		\begin{tabular}{|c|c|c|c||c|c|c|c||c|c|c|c|c|c|c|c|c} \hline
$M_{1/2}$&$M_{\rm V}$&$\tan\beta$&$m_{t}$&$\Omega h^2$&$\langle \sigma v \rangle _{f \overline{f}} $&$\langle \sigma v \rangle _{\gamma \gamma} $&$\langle \sigma v \rangle _{\gamma Z} $&$m_{\chi^0_1}$&$m_{\widetilde{\tau}_{1}}$&$\Delta M$&$m_{\chi^{0}_{2},\chi^{\pm}_{1}}$&$m_{\widetilde{t}_{1}}$&$m_{\widetilde{g}}$&$m_{\widetilde{u}_{R}}$&$m_h$ \\ \hline \hline	
$	775	$&$	4800 $&$	22.5	$&$	174.4	$&$	0.122	$&$	68.5 \times 10^{-30}	$&$	2.61 \times 10^{-30}	$&$	0.98 \times 10^{-30}	$&$	161	$&$	169	$&$	7.87	$&$	342	$&$	861	$&$1047$&$1475$&$124.4$	\\	\hline
$	774	$&$	4821	$&$	23.0	$&$	174.4	$&$0.036$&$	77.9 \times 10^{-30}	$&$3.01 \times 10^{-30}	$&$1.11 \times 10^{-30}	$&$	160	$&$	162	$&$	1.95	$&$	341	$&$	860	$&$	1046	$&$	1473	$&$124.4$	\\	\hline
$	774	$&$	4851	$&$	23.1	$&$	174.4	$&$	0.020	$&$	81.1 \times 10^{-30}	$&$	3.19 \times 10^{-30}		$&$	1.16 \times 10^{-30}		$&$	160	$&$	160	$&$0.06		$&$	341	$&$	860	$&$1046$&$1473$&$124.4$	\\	\hline\hline
$	990	$&$	8044	$&$	23.3	$&$	174.4	$&$	0.120	$&$	45.6 \times 10^{-30}	$&$	1.73 \times 10^{-30}		$&$	0.69 \times 10^{-30}		$&$	214	$&$	220	$&$	6.43	$&$	449	$&$	1104	$&$1328$&$1824$&$125.1$	\\	\hline
$	990	$&$	8070	$&$	23.6	$&$	174.4	$&$	0.056	$&$	47.6 \times 10^{-30}	$&$	1.88 \times 10^{-30}		$&$	0.76 \times 10^{-30}		$&$	213	$&$	216	$&$ 2.06		$&$	449	$&$	1104	$&$1328$&$1824$&$125.1$	\\	\hline
$	1000	$&$	8083	$&$	23.7	$&$	174.4	$&$	0.036	$&$	46.7 \times 10^{-30}	$&$1.93 \times 10^{-30}			$&$	0.77 \times 10^{-30}		$&$	216	$&$	216	$&$	0.21	$&$	454	$&$	1116	$&$1341$&$1841$&$125.2$	\\	\hline\hline
$	1200	$&$	30,830	$&$	24.3	$&$	173.3	$&$	0.122	$&$	20.5 \times 10^{-30}	$&$	1.26 \times 10^{-30}		$&$	0.51 \times 10^{-30}		$&$	276	$&$	281	$&$	4.54	$&$	572	$&$	1335	$&$1633$&$2102$&$124.1$	\\	\hline
$	1200	$&$	30,830	$&$	24.4	$&$	173.3	$&$	0.084	$&$	20.8 \times 10^{-30}	$&$	1.31 \times 10^{-30}		$&$0.53 \times 10^{-30}			$&$	276	$&$	279	$&$	2.07	$&$	572	$&$	1335	$&$1634$&$2102$&$124.1$	\\	\hline
$	1200	$&$	30,830	$&$	24.5	$&$	173.3	$&$	0.056	$&$	21.1 \times 10^{-30}	$&$	1.36 \times 10^{-30}		$&$	0.55 \times 10^{-30}		$&$	276	$&$	277	$&$	0.08	$&$	572	$&$	1335	$&$1634$&$2102$&$124.1$	\\	\hline\hline
$	1500	$&$	27,636	$&$	24.7	$&$	174.4	$&$	0.122	$&$	8.88 \times 10^{-30}	$&$	0.85 \times 10^{-30}		$&$	0.35 \times 10^{-30}		$&$	349	$&$	351	$&$	2.06	$&$	717	$&$	1661	$&$2009$&$2602$&$126.3$	\\	\hline
$	1500	$&$	27,636	$&$	24.8	$&$	174.4	$&$	0.086	$&$	8.93 \times 10^{-30}	$&$	0.88 \times 10^{-30}		$&$0.36 \times 10^{-30}		$&$	349	$&$	349	$&$	0.07	$&$	717	$&$	1661	$&$2009$&$2602$&$126.3$	\\	\hline
		\end{tabular}
		\label{tab:benchmarks}
\end{table*}

\begin{table*}[htp]
	\centering
	\caption{The ten No-Scale \fsu5 benchmarks of Table~\ref{tab:benchmarks}, with the IB photon
flux $\Phi _{IB}$ from $\widetilde{\chi} \widetilde{\chi} \to f \overline{f} \gamma$ events, the photon
flux $\Phi _{\gamma \gamma}$ from $\widetilde{\chi} \widetilde{\chi} \to \gamma \gamma$ events, and the photon
flux $\Phi _{\gamma Z}$ from $\widetilde{\chi} \widetilde{\chi} \to \gamma Z$ events. The IB flux has been
integrated across energy relative to the differential flux plotted in Figure~\ref{fig:ibyield}.
All fluxes are also integrated over the solid line-of-sight angle from the center of our galaxy, taking
a detector acceptance of 2.5 steradians corresponding to the LAT instrument's 20\% sky field of view,
and are in units of photons ${\rm cm^{-2}~sec^{-1}}$.
All fluxes are calculated with {\tt DarkSUSY~5.1.1}
The $\gamma \gamma$ flux includes the factor of 2 for the two photons.
For the local dark matter relic density, we use the value $\rho_0 = 0.3$ GeV/${\rm cm^3}$, with the
spherically symmetric NFW halo profile.
The column entry $\Phi _{IB}/\Phi _{\gamma \gamma}$ is indicative of the increase in
the magnitude of the IB flux over the gamma pair flux, and the adjacent column
$\Phi _{\gamma \gamma}/ \Phi _{\gamma Z}$ likewise compares the gamma pair flux to that of the photon plus Z-boson.
The final two columns provide the gamma radiation energy in GeV at the IB spectrum peak
and its relation to the LSP mass in GeV.}
		\begin{tabular}{|c|c|c|c||c|c|c|c|c||c|c|} \hline
$M_{1/2}$&$M_{\rm V}$&$\tan\beta$&$m_{t}$&$\Phi _{IB}$&$\Phi _{\gamma \gamma} $&$\Phi _{\gamma Z} $&$\Phi _{IB}/\Phi _{\gamma \gamma}$&$\Phi _{\gamma \gamma}/\Phi _{\gamma Z}$& ${\rm IB~Peak}$&$m_{\chi^0_1}$ \\ \hline \hline
$	775	$&$	4800 $&$	22.53	$&$	174.4	$&$	3.9 \times 10^{-12}	$&$	 4.9\times 10^{-13}	$&$	 5.5\times 10^{-14}	$&$8.1$ & $8.7$& $148$&$161$\\	\hline
$	774	$&$	4821	$&$	22.95	$&$	174.4	$&$	5.8 \times 10^{-12}	$&$	 5.6\times 10^{-13}	$&$	 6.2\times 10^{-14}	$&$10.3$ & $9.0$& $153$&$160$\\	\hline
$	774	$&$	4851	$&$	23.08	$&$	174.4	$&$	4.6 \times 10^{-12}	$&$	 4.0\times 10^{-13}	$&$	 4.4\times 10^{-14}	$&$11.5$ & $9.1$& $156$&$160$\\	\hline\hline
$	990	$&$	8044	$&$	23.34	$&$	174.4	$&$	1.5 \times 10^{-12}	$&$	 1.8\times 10^{-13}	$&$	 2.3\times 10^{-14}	$&$8.4$ & $8.0$& $200$&$214$\\	\hline
$	990	$&$	8070	$&$	23.61	$&$	174.4	$&$	1.9 \times 10^{-12}	$&$	 2.0\times 10^{-13}	$&$	 2.4\times 10^{-14}	$&$9.8$ & $8.1$& $204$&$213$\\	\hline
$	1000	$&$	8083	$&$	23.73	$&$	174.4	$&$	2.1 \times 10^{-12}	$&$	 2.0\times 10^{-13}	$&$	 2.4\times 10^{-14}	$&$10.7$ & $8.1$& $211$&$216$\\	\hline\hline
$	1200	$&$	30,830	$&$	24.26	$&$	173.3	$&$	6.5 \times 10^{-13}	$&$	 7.9\times 10^{-14}	$&$	 1.0\times 10^{-14}	$&$8.2$ & $7.8$& $262$&$276$\\	\hline
$	1200	$&$	30,830	$&$	24.41	$&$	173.3	$&$	7.3 \times 10^{-13}	$&$	 8.2\times 10^{-14}	$&$	 1.1\times 10^{-14}	$&$8.9$ & $7.8$& $266$&$276$\\	\hline
$	1200	$&$	30,830	$&$	24.53	$&$	173.3	$&$	8.1 \times 10^{-13}	$&$	 8.5\times 10^{-14}	$&$	 1.1\times 10^{-14}	$&$9.6$ & $7.8$& $271$&$276$\\	\hline\hline
$	1500	$&$	27,636	$&$	24.67	$&$	174.4	$&$	3.0 \times 10^{-13}	$&$	 3.4\times 10^{-14}	$&$	 4.3\times 10^{-15}	$&$8.8$ & $7.8$& $336$&$349$\\	\hline
$	1500	$&$	27,636	$&$	24.77	$&$	174.4	$&$	3.3 \times 10^{-13}	$&$	 3.5\times 10^{-14}	$&$	 4.4\times 10^{-15}	$&$9.4$ & $7.8$& $343$&$349$\\	\hline
		\end{tabular}
		\label{tab:flux}
\end{table*}

It is well known that the annihilation cross section of the LSP neutralinos into a pair of light SM fermions
is strongly suppressed by a factor $m_f^2/m_{\chi_1^0}^2$ due to the helicity properties of a highly
non-relativistic pair of Majorana neutralinos. However, such suppression can be evaded if the
fermion final states contain an additional photon $f {\bar f} \gamma$, particularly when the photon is
emitted from virtual sfermions with a mass close to the LSP neutralino. Therefore, the IB effects may
explain the 133 GeV Fermi-LAT gamma ray line~\cite{Bringmann:2012ez, Shakya:2012fj}, or may predict a
higher energy (for example 200 GeV) gamma ray line in No-Scale \fsu5. Furthermore, the EW or strong gauge
boson IBs have considerably larger rates due to the larger gauge coupling constants.  Recently, a complete
calculation of the leading EW corrections to the LSP neutralino annihilations for various final
states~\cite{Bringmann:2013oja} shows that such corrections may significantly enhance the
annihilation rates. Although those processes do not generate the pronounced spectral features in gamma
rays like the corresponding electromagnetic (EM) corrections, the integrated photon yield may be
enhanced up to two orders of magnitude compared to the tree level results, which may also be probed by the
ongoing Fermi Space Telescope experiment. As such, we have ample motivation to study those regions of the viable
parameter space with small mass differences between the LSP neutralino and light stau.

Our mission here then is to augment the SUSY neutralino annihilation rates to enhance detection
opportunity for a nearly pure bino LSP ($> 99\%$ bino). Through near degeneracy amongst the lightest
slepton and light bino masses, we can certainly increase the annihilation rate and boost IB effects to a
dominant contribution, albeit with downward pressure on the bino relic density. For a SUSY bino, this
requires a compressed $\Delta M(\widetilde{\chi}_1^0,\widetilde{\tau}_1) \simeq 0-2$ GeV, with associated decays
proceeding through an off-shell or on-shell tau accordingly.
Compression of the light stau mass to $\Delta M(\widetilde{\chi}_1^0,\widetilde{\tau}_1) \simeq 0-2$
GeV can be achieved in No-Scale \fsu5 quite naturally via slight shifts of the low energy boundary
condition $\tan\beta$. The resultant minor increase in $\tan\beta$ does lead to marginally enhanced
light stau mixing effects in the stau sector, slightly lowering the light stau mass. Satisfaction
of the CDM relic density in a traditional thermal manner leads to an intrinsic escalation in the
baseline value of this parameter, from $\tan\beta \simeq 19.5$ to $\tan\beta \simeq 25$
for a corresponding upward escalation in the gaugino mass from $M_{1/2} \simeq 400$ to $M_{1/2} \simeq
1500$~\cite{Li:2013naa}. Because of this, the supplemental incrementation of $\tan\beta$ required to
squeeze the light stau mass and LSP to near degeneracy recedes with an inflating SUSY mass scale. The
positive deviation in $\tan\beta$, with possibly small shifts in the gaugino mass $M_{1/2}$ and flippon
mass $M_V$, are all that are required to achieve the 0--2 GeV delta between the light stau mass and the LSP in
the large unprobed region of the parameter space. In particular, no variation of the top quark mass $m_t$
(within its experimental uncertainty) is necessary. As a result, the SUSY spectrum undergoes only a
negligible transition, and thus the rich phenomenology (setting aside the relic density constraint,
which must now be satisfied through non-thermal mechanisms) prevails wholly preserved. Indeed, the
wedge of model space remains relatively static and persists in the form of Ref.~\cite{Li:2013naa}, the
lone exception being small shifts in the $\tan\beta$ contours and indiscernible shifts in $M_{1/2}$ and
$M_{V}$.

\begin{figure*}[htp]
        \centering
        \includegraphics[width=0.75\textwidth]{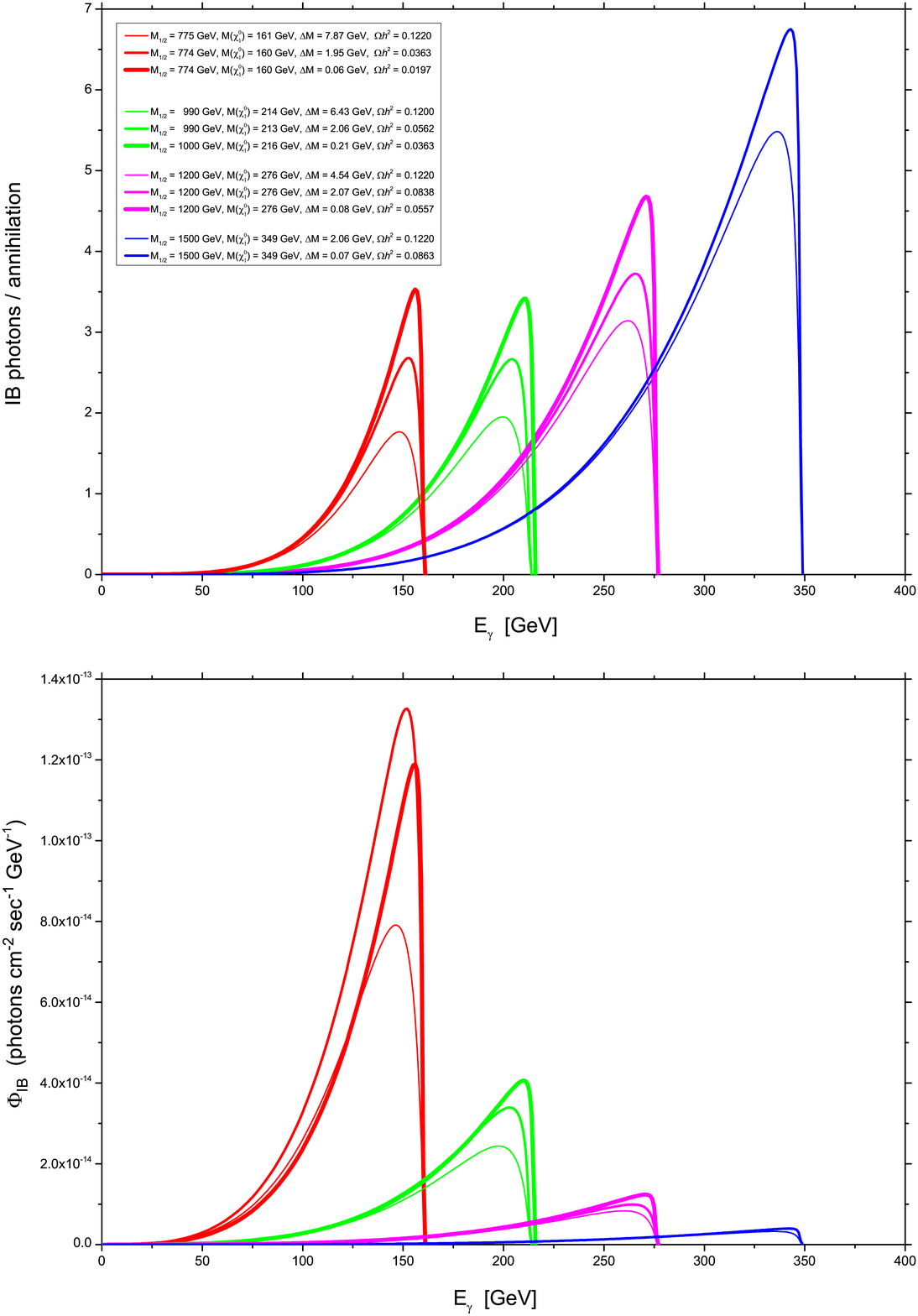}
        \caption{No-Scale \fsu5 Electromagnetic IB spectrum, given in terms of photons per annihilation (top frame) and
differential flux (bottom frame), as a function of energy. All curves represent the benchmarks given in Tables~\ref{tab:benchmarks}-\ref{tab:flux}.
The thin curves (lower) in both frames satisfy the Planck satellite CDM relic density measurements $\Omega h^2 = 0.1199 \pm 0.0027$,
while the thicker curves (middle) possess $\Delta M(\widetilde{\chi}_1^0,\widetilde{\tau}_1) \simeq 2$ GeV and the thickest
curves (upper) have $\Delta M(\widetilde{\chi}_1^0,\widetilde{\tau}_1) \simeq 0$ GeV. Inclusion of the EW
IB photon flux enhancement is reserved for a future work.  The $M_{1/2} = 774-775$ GeV benchmarks are
consistent with the previously observed 133 GeV monochromatic gamma ray line. The $\Delta M$ value given
in the plot legend refers to the lightest neutralino and light stau mass difference. All IB photon counts
and fluxes are calculated with {\tt DarkSUSY~5.1.1}. For the local dark matter relic density, we use the
value $\rho_0 = 0.3$ GeV/${\rm cm^3}$.  All differential fluxes are in units of photons ${\rm cm^{-2}~sec^{-1}~GeV^{-1}}$ and all
masses are in GeV. The $\Omega h^2$ shown in the plot legend is the thermal neutralino relic density
calculated with {\tt MicrOMEGAs~2.4}. For those benchmarks with $\Omega h^2 < 0.1199 \pm 0.0027$, the
Planck satellite measured relic density can be generated via non-thermal mechanisms. The curves
demonstrate that compression of the lightest bino neutralino and light stau mass delta does in fact
enhance the EM IB effects.}
        \label{fig:ibyield}
\end{figure*}

From this perspective, the No-Scale \fsu5 SUSY spectra corresponding to the wedge of viable model space
provided in Ref.~\cite{Li:2013naa}, duly suppressing the light stau mass, are potentially testable by the Fermi Space Telescope 
or a future gamma ray telescope; moreover, the two variations in determination of the light stau mass may be observationally
distinguished. Crucially, experimental results from both the LHC and the LAT can be connected to the
same SUSY spectrum, providing the type of cross-correlation testing which may play a significant role in
substantiating any SUSY GUT model. In particular, probing of a specific
$(\widetilde{\chi}_1^0, \widetilde{t}_1, \widetilde{g}, \widetilde{q})$ point in the SUSY parameter space
may potentially be achieved via dual experimental methodologies. This is possible since the No-Scale
\fsu5 SUSY spectrum exhibits the rather special attribute of leading order $en~ masse$ proportionality
to only $M_{1/2}$. Specifically, the internal physics of \fsu5 are predominantly invariant under a
numerical rescaling of only $M_{1/2}$. Consequently, each sparticle within the SUSY spectrum can be
multiplicatively adjusted by an identical trivial rescaling of only $M_{1/2}$, though the linear slope
relationship between $M_{1/2}$ and each sparticle can vary. From a practical point of view, this property
of No-Scale \fsu5 permits the SUSY spectrum to be approximately determined from only a given value of
$M_{1/2}$, or alternatively, from only a given value of any other sparticle mass, exhibiting the
pragmatic predictive elegance of the model.

The final ingredient of our strategy involves derivation of a suitable set of benchmarks for comparison to
experiment. We present ten benchmarks in Table~\ref{tab:benchmarks}, with gaugino mass $M_{1/2}$,
flippon mass $M_V$, $\tan\beta$, top quark mass $m_t$, relic density $\Omega h^2$, EM $f \overline{f}$,
$\gamma \gamma$, and $\gamma Z$ annihilation cross-sections, SUSY masses, and light Higgs boson
mass. All benchmark LSP compositions are greater than 99\% bino. The points have been extracted from a
broad numerical scan, utilizing {\tt MicrOMEGAs~2.1}~\cite{Belanger:2008sj} to compute SUSY mass
spectra and a proprietary modification of the {\tt SuSpect~2.34}~\cite{Djouadi:2002ze} codebase to
run the flippon enhanced RGEs. To be consistent with previous No-Scale \fsu5 parameter space
analyses~\cite{Li:2011xu,Li:2013naa}, we show in Table~\ref{tab:benchmarks} the thermal relic
density as computed by the updated routines in {\tt MicrOMEGAs~2.4}~\cite{Belanger:2010gh}. Serving as a
secondary verification, we further compute the thermal relic density with {\tt
DarkSUSY~5.1.1}~\cite{Gondolo:2004sc,DarkSUSY}, reading as input an
SLHA~\cite{Skands:2003cj,Allanach:2008qq} mass file generated from the flippon enhanced RGEs in our
proprietary version of the {\tt SuSpect~2.34}~\cite{Djouadi:2002ze} codebase, finding only a
small variation in the respective relic density computations. The annihilation cross-sections
$\langle \sigma v \rangle _{f \overline{f}}$, $\langle \sigma v \rangle _{\gamma \gamma}$, and
$\langle \sigma v \rangle _{\gamma Z}$ are calculated with both {\tt MicrOMEGAs~2.4} and {\tt DarkSUSY~5.1.1},
where we show the average of the two calculations in Table~\ref{tab:benchmarks}.
The total $\langle \sigma v \rangle _{f \overline{f}}$ annihilation cross-section includes the only
three non-negligible contributions in No-Scale \fsu5 for a nearly pure SUSY bino:
$\langle \sigma v \rangle _{f \overline{f}} = \langle \sigma v \rangle _{\tau^+ \tau^-} +
\langle \sigma v \rangle _{t\overline{t}} + \langle \sigma v \rangle _{b \overline{b}}$.
The $\Delta M$ value in Table~\ref{tab:benchmarks} refers specifically to the light neutralino and light stau mass
difference, which we are compressing to increase the annihilation rate and IB effects. The light
Higgs boson mass $m_h$ in Table~\ref{tab:benchmarks} includes both the
tree level+1-loop+2-loop+3-loop+4-loop contributions and the additional vector-like flippon contribution~\cite{Li:2013naa}.

Expected photon flux rates are listed in Table~\ref{tab:flux} for the annihilation channels $\widetilde{\chi} \widetilde{\chi}
\to f \overline{f} \gamma$, $\widetilde{\chi} \widetilde{\chi} \to \gamma \gamma$, and
$\widetilde{\chi} \widetilde{\chi} \to \gamma Z$, for the same ten No-Scale \fsu5 benchmarks of
Table~\ref{tab:benchmarks}. For the local dark matter relic density, we use the value $\rho_0 = 0.3$ GeV/${\rm cm^3}$,
adopting the spherically symmetric NFW dark matter halo profile.
The square of the dark matter density is integrated along the line of sight for each orientation within an angular
detector acceptance of 2.5 steradians (sr) about the galactic center. This value is selected in correspondence with the
LAT instrument's field of view, which encompasses about 20\% of the sky at any given moment.  Results are not overly sensitive
to this parameter, given a value sufficiently wide to encapsulate the region of primary density.
Since the IB scenario represents a continuum of radiation frequencies, the differential fluxes plotted in the lower panel of Figure~\ref{fig:ibyield}
are integrated across energy to yield consistent units of photon counts per square centimeter per second in Table~\ref{tab:flux}. 
All fluxes are computed with {\tt DarkSUSY~5.1.1}.
The $\gamma \gamma$ flux includes the factor of 2 for the two photons. The ratio $\Phi _{IB}/\Phi _{\gamma \gamma}$ in
Table~\ref{tab:flux} represents the magnitude of the integrated IB flux relative to the $\gamma \gamma$ line flux, which
provides an advantage of about 10 across the full model space.  Likewise, the column $\Phi _{\gamma \gamma} / \Phi _{\gamma Z}$.
reports the ratio of monochromatic flux rates for a gamma pair relative to a gamma plus Z-boson, which similarly 
yields an advantage of one magnitude order across the model space.

It is evident from Figure~\ref{fig:ibyield} that compressing the light bino neutralino and light
stau does indeed enhance the EM IB effects for the benchmarks of Table~\ref{tab:benchmarks}. The curves
in the top frame of Figure~\ref{fig:ibyield} depict the number of IB photons per annihilation resulting
from annihilation into charged particles. The bottom frame illustrates the IB flux $\Phi_{IB}$ energy
spectrum for the same ten benchmarks. The thin curves (lower) in both frames represent a region of the
No-Scale \fsu5 model space where the thermal LSP relic density can satisfy the Planck satellite CDM
measurements $\Omega h^2 = 0.1199 \pm 0.0027$~\cite{Ade:2013zuv}. The thicker curves (middle) in both
frames possess an LSP and light stau mass difference of about 2 GeV, with the thickest curves (upper) having
a degenerate LSP and light stau, with possibly a long-lived light stau in this degenerate scenario. All IB
photon counts in Figure~\ref{fig:ibyield} are computed with {\tt DarkSUSY~5.1.1}, as are the IB fluxes. Clearly, the
EM IB photon count, and hence the flux, increases for smaller $\Delta M$, an effect we presume will be
enhanced when also including the EW contributions~\cite{Bringmann:2013oja}. We leave the numerical
results of the EW IB photon yield and additional flux for a future work~\cite{LMNW-P}. At this juncture, we
are content with a projection that the photon counts and fluxes in Figure~\ref{fig:ibyield} could be
amplified via the additional EW IB contributions~\cite{Bringmann:2013oja}.

Our scale for the benchmarks in Tables~\ref{tab:benchmarks}-\ref{tab:flux} and
Figure~\ref{fig:ibyield} begins at $M_{1/2} = 775$ GeV, which is in the vicinity of the scale threshold
that may be considered firmly excluded
from the No-Scale \fsu5 model space by the LHC SUSY search, as based upon a Monte Carlo event
analysis~\cite{Li:2013hpa}. We select sufficient points to provide thorough coverage of the entire
viable model space.  We direct attention to the region of the parameter
space exemplified by the $M_{1/2} = 774-775$ GeV benchmarks of
Tables~\ref{tab:benchmarks}-\ref{tab:flux} as that consistent with an upper 2$\sigma$ limit on the
WIMP mass that can explain the previously observed 133 GeV monochromatic gamma ray line.
Comparing Figure~\ref{fig:ibyield} with Table~\ref{tab:flux}, it is apparent that
compression of the $\Delta M(\widetilde{\chi}_1^0,\widetilde{\tau}_1)$ mass gap substantially
strengthens the IB signal in the narrowly peaked spectral range close to the LSP mass, whereas
the advantage in integrated photon flux is less pronounced; this is relevant given higher experimental
sensitivity to signals that more closely approximate a line spike.

\section{Summary of Experimental Prospects}

In this final section, we attempt to make a quantitative, if in some regards na\"{\i}ve,
assessment of the experimental prospects of the various \fsu5 model benchmarks previously described.
The primary metric for assessment will be the integrated photon flux, {\it i.e.} the area under each
differential flux curve displayed in the lower element of Figure~\ref{fig:ibyield}, in units of
photons ${\rm cm^{-2}~sec^{-1}}$, as reported in Table~\ref{tab:flux}.
Since both background (following a power law with spectral index $-2$) and the internal bremsstrahlung
signal accrue in linear proportion with time, the $S/\sqrt{B}$
signal to background discriminant may be expected to scale as the square root of time.
Based upon four years of data collection in whole-sky survey mode (achieving a full $4\pi$ steradian coverage
once per two earth orbits), the Fermi collaboration has established sensitivity at five standard deviations
to gamma flux rates above about $3-4 \times 10^{-9}~{\rm cm^{-2}~sec^{-1}}$ for line sources positioned at
high galactic latitudes~\cite{LATsensitivity}; the sensitivity is diminished by about half an order of magnitude in the highly
active galactic center. Taking an active Fermi mission lifetime of ten years, one sees that the data doubling
advantage has already been largely depleted in the existing results, although the remaining multiple of 2.5
in integrated time may yet garner an improvement of around 1.6 deviations in sensitivity; in other words,
any potential discovery apparent by the end of the Fermi mission should already be showing evidence above
three standard deviations.  Likewise, the expected end of mission line sensitivity may be projected at
about $2 \times 10^{-9}~{\rm cm^{-2}~sec^{-1}}$.

The root-$t$ scaling is actually a bit pessimistic for signals approximating a line width, and better sensitivity is possible.
Additionally, the Fermi instrument has begun a transition toward more targeted observation of the galactic center for the remainder of its mission, which may
garner an additional factor of about two in sensitivity, admitting however that baseline sensitivities
are lower in this region.  Likewise, substantial improvements in understanding of the detector and relevant analysis techniques
are poised to reduce background contamination and improve overall instrument sensitivity~\cite{Atwood:2013rka}; we
likewise assign a factor of about two to processing upgrades of this type, which are retroactive to already collected data.
Holding backgrounds constant, a further reduction in the signal flux by a factor
around $3/5$ would still be capable of presenting strong evidence for a scale-localized excess.  Together, then,
we set a working threshold around $3 \times 10^{-10}~{\rm cm^{-2}~sec^{-1}}$ on any potentially visible
gamma flux. Given continuum dispersion of the IB gamma signal, it is somewhat over optimistic to apply
sensitivities extrapolated from line-signal searches, and this deficiency becomes more pronounced at
higher mass scales with widening and flattening of the signal profile, as is visible in Figure~\ref{fig:ibyield}.
Nevertheless, it is important to recognize that any IB gamma signal may be compounded with line signals
from loop order neutralino annihilation to gamma pairs and/or gamma Z, in the same basic spectral range,
although potentially substantially suppressed, as indicated in Table~\ref{tab:flux}.  Without an appreciable
boost factor $\mathcal{O}$(50--100) in the computed annihilation rate, the \fsu5 IB gamma flux, while more favorable for detection
than the flux associated with mono-energetic line sources, may remain obscured by background processes to the
LAT instrument.  However, if there is any validity to the existing 130~GeV signal, then it becomes quite likely
that some undiagnosed boost factor is actually in play.  Plausible sources of this upward shift in the flux include
underestimation of the local dark matter density (or corrections to the assumption of a smooth profile distribution),
and internal bremsstrahlung contributions from EW or strong gauge bosons. 

As a closing note, we draw attention to the increase in the thermally
produced bino relic density in Table~\ref{tab:benchmarks} for those points with $\Delta
M(\widetilde{\chi}_1^0,\widetilde{\tau}_1) \simeq 0-2$ GeV,
as the gaugino mass $M_{1/2}$ is lifted; this is due primarily to the incrementally larger LSP mass, and a corresponding
slow increase in the value of $\tan\beta$, which tracks the elevation in $M_{1/2}$, automatically enhancing the light
stau mixing for larger SUSY mass scales. Interestingly, the viable No-Scale \fsu5 parameter space terminates near $M_{1/2} \sim 1500$ GeV,
with a nearly degenerate light stau and LSP, while concurrently maintaining the Planck observed relic density.
Furthermore, if we consider an off-shell tau, the parameter space can be extended up to $M_{1/2} \sim 1700$
GeV before incurring a charged light stau as the LSP. In this uppermost region of the model space, no alternate
measures, such as non-thermally produced WIMPS, need be invoked to generate the correct relic density.
This very large $M_{1/2} \sim 1500$ GeV region may be probed by future gamma ray probe experiments, and any possible
gamma ray line signals could be directly correlated to LHC results, where, given the strong light stau and LSP
neutralino mass degeneracy in this portion of the model, one may make an additional intriguing prediction
for LHC phenomenology: in light stau production, the tau and LSP neutralino missing momentum signal will be collinear.

\section{Conclusions}

We presented here a methodology for testing No-Scale Supergravity with the FERMI satellite's Large Area Telescope,
and similar future gamma ray telescopes.
For our testing vehicle, we chose the supersymmetric grand unified model No-Scale Flipped
$SU(5)$ with extra vector-like flippon multiplets derived from F-Theory, dubbed \fsu5. Building upon ample
extant phenomenological motivation for No-Scale \fsu5, we discussed the potentially significant
empirical support recently provided to cosmological models of inflation based upon No-Scale Supergravity by
intrinsic Starobinsky-like conformance with the Planck measurements, for a suitable choice of superpotential parameters.
Given this impetus, we discussed how compressing the light stau and LSP mass
difference can increase the internal bremsstrahlung effects and thus enhance the photon count from
annihilation to elevate detection probabilities, albeit with a reduced bino relic density.
We additionally explained how the Planck satellite observed relic density can nevertheless be generated through
a non-thermal mechanism. For concrete examples, we gave several benchmark points with light stau and LSP
mass differences of 0--2 GeV, achieved by slight upward shifts in the low energy boundary condition
$\tan\beta$, in conjunction with negligible variations in the gaugino mass $M_{1/2}$ and flippon mass
$M_{V}$; these modifications leave the SUSY spectrum, aside from the light stau mass, unchanged, preserving the rich
phenomenology (modulo appeal to non-thermal mechanisms of relic density generation) that is currently
being probed by the LHC and several other Beyond the Standard Model (BSM) experiments.
While the IB mechanism emerges as a more favorable context for observing a gamma ray signal generated
consistently with the \fsu5 model than monochromatic sources, a clear signal in the present generation
instrument still requires a boost of order $\mathcal{O}$(50--100) in the expected rate of flux.


\begin{acknowledgments}
This research was supported in part by the DOE grant DE-FG03-95-Er-40917 (DVN) and by the Natural Science
Foundation of China under grant numbers 10821504, 11075194, 11135003, and 11275246 (TL).
We also thank Sam Houston State University for providing high performance computing resources.
\end{acknowledgments}


\bibliography{bibliography}

\end{document}